\documentclass[aps,prl,twocolumn]{revtex4-2}
\usepackage{setspace}
\usepackage{amsmath}
\usepackage{siunitx}
\usepackage{amsthm,amsmath,amssymb}
\usepackage{mathrsfs}
\usepackage{bm}
\usepackage{graphicx}
\usepackage[nearskip,margin = 0pt]{subfig}
\usepackage{verbatim}
\usepackage{amsfonts}
\usepackage{esint}
\usepackage[colorlinks,linkcolor=blue,anchorcolor=blue,citecolor=blue,urlcolor=black]{hyperref}
\usepackage{float}
\usepackage{soul}
\usepackage{natbib}

\begin{document}
\title{Single-particle vibrational spectroscopy using optical microresonators}
\author{Shui-Jing Tang$^{1}$, Mingjie Zhang$^{1}$, Jialve Sun$^{2}$, Jia-Wei Meng$^{1}$, Xiao Xiong$^{1}$, Qihuang Gong$^{1,3}$, Dayong Jin$^{4}$,  Qi-Fan Yang$^{1,3}$, Yun-Feng Xiao$^{1,2,3\ast}$ \\
\vspace{3pt}
$^1$Frontiers Science Center for Nano-optoelectronics and State Key Laboratory for Mesoscopic Physics, School of Physics, Peking University, Beijing 100871, China\\
$^2$National Biomedical Imaging Center, Peking University, Beijing 100871, China\\
$^3$Collaborative Innovation Center of Extreme Optics, Shanxi University, Taiyuan 030006, China\\
$^4$Institute for Biomedical Materials and Devices (IBMD), Faculty of Science, University of Technology Sydney, Sydney, New South Wales, Australia\\
\vspace{3pt}
$^\ast$Corresponding author: yfxiao@pku.edu.cn}

\begin{abstract}
Vibrational spectroscopy is a ubiquitous technology that derives the species, constituents, and morphology of an object from its natural vibrations. 
However, the vibrational spectra of mesoscopic particles – including most biological cells - have remained hidden from existing technologies.
These particles are expected to vibrate faintly at megahertz to gigahertz rates, imposing unpractical sensitivity and resolution for current optical and piezoelectric spectroscopy.
Here we demonstrate the real-time measurement of natural vibrations of single mesoscopic particles using an optical microresonator,
extending the reach of vibrational
spectroscopy to a new spectral window.
Conceptually, a spectrum of vibrational modes of the particles is stimulated photoacoustically, and correlated to a high-quality-factor optical resonance for the ultrasensitive readout.
Experimentally, this scheme is testified by measuring mesoscopic particles with different constituents, sizes, and internal structures, showing an unprecedented signal-to-noise ratio of 50 dB and detection bandwidth over 1 GHz.
This new technology is further applied for the biomechanical fingerprinting of single microbial cells with different species and living states.
The present method opens up new avenues to study single-particle mechanical properties in vibrational degrees of freedom, and may find applications in photoacoustic sensing and imaging, cavity optomechanics and biomechanics.
\end{abstract}
\maketitle

\captionsetup[figure]{labelfont={bf},justification=raggedright}

\begin{figure*}[ht]
\centering
\includegraphics[width=15cm]{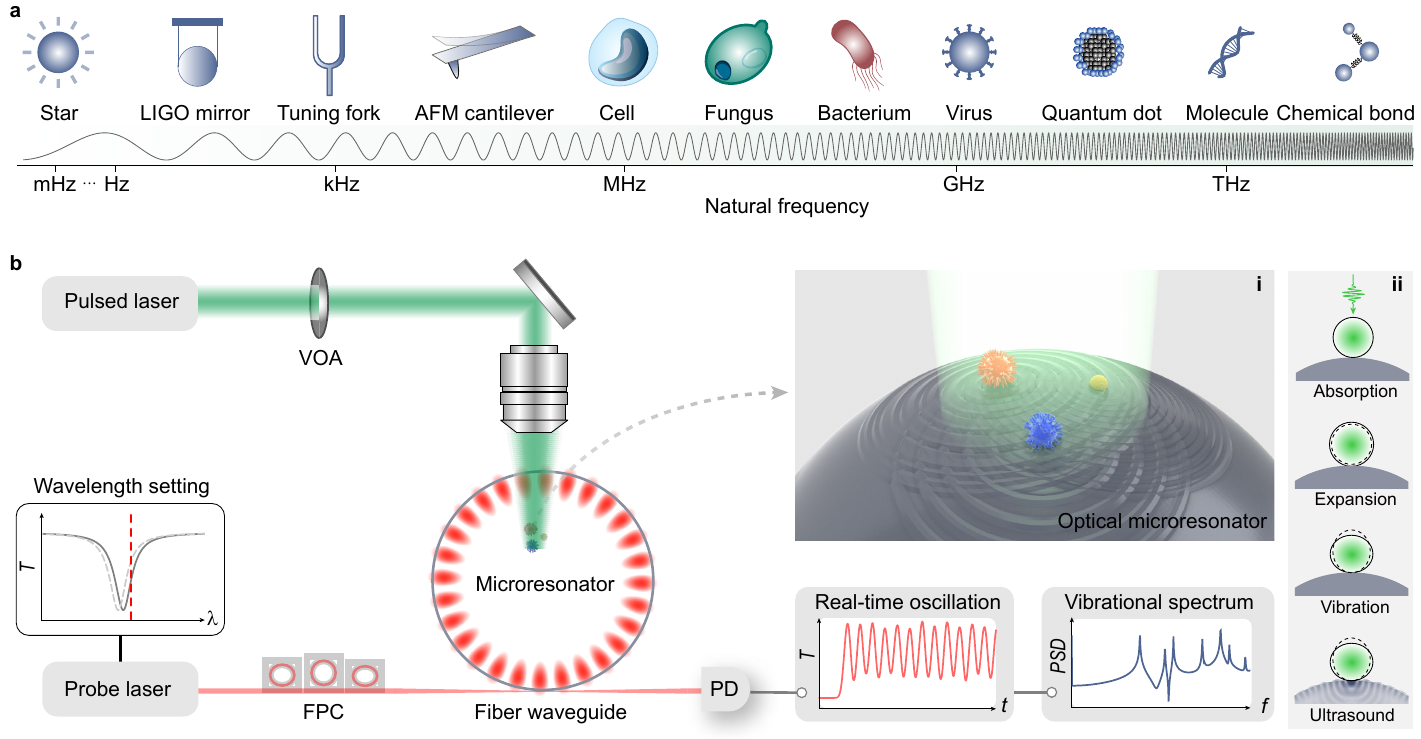}
\caption{\textbf{Microresonator-based vibrational spectroscopy. a}, Frequencies of natural vibrations for objects with different size scales. 
{\bf b}, Experimental apparatus. Particles deposited on the microresonator are irradiated by a pulsed laser to stimulate their natural vibrations. 
A continuous-wave probe laser is used to excite the optical whispering-gallery mode by setting the wavelength slightly detuned from the resonance, and the optical transmission is recorded by a photodetector.
Inset i: the enlarged view of vibrating particles on the optical microresonator. Inset ii: photoacoustic excitation of natural vibrations and their acoustic coupling to the optical mode (from top to bottom).
PSD, power spectral density; 
VOA, variable optical attenuator; FPC, fiber polarization controller; PD, photodetector.
}
\label{fig1}
\end{figure*}

\begin{figure*}[ht]
\centering
\includegraphics[width=18cm]{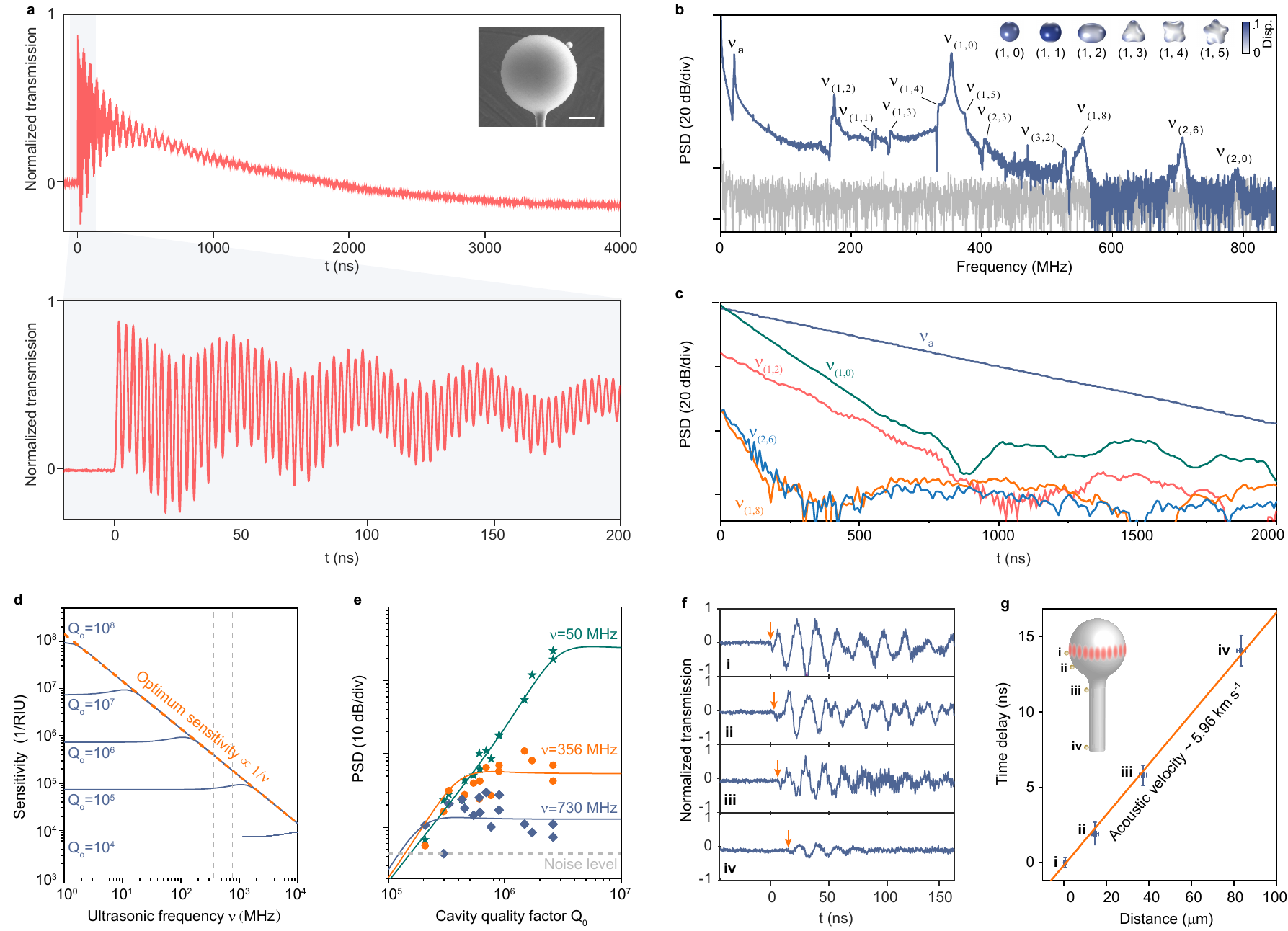}
\caption{\textbf{Temporal and spectral measurement of natural vibrations of a single particle. a}, Transmitted power of the probe laser when a single polystyrene particle deposited on the microresonator is radiated by the pulsed laser. Inset: scanning-electron-microscope (SEM) image.
Scale bar: 20 \textmu m. \textbf{b}, The vibrational spectrum (navy) of the particle derived from the trace in {\bf a}. 
The measurement noise floor is also plotted (grey). Insets: the displacement patterns of a few vibrational modes of a free sphere. \textbf{c}, Peak intensities of the vibrational modes as a function of time. 
\textbf{d}, Theoretical sensitivity of the microresonator sensor depending on both $Q_{\text{o}}$ and ultrasonic frequency $\nu$. 
{\bf e}, $Q$-dependent optical response to particle's vibrations at three frequencies (grey dashed lines in \textbf{d}). Symbols: measured peak intensities extracted from vibrational spectra of a 1.4-\textmu m-radius polystyrene sphere. Solid curves: theoretical results. 
\textbf{f}, The transmissions of the probe laser when the particle placed on different locations (the inset in \textbf{g}) are irradiated at $t=0$.
The arrows indicate the time when the laser starts to be modulated. 
\textbf{g}, The response time as a function of the distance between the optical mode and the particles.  Error bars indicate the uncertainties of the measured time delays and propagating distances of acoustic waves.  
}
\label{Fig2}
\end{figure*}

\begin{figure*}[ht]
\centering
\includegraphics[width=\linewidth]{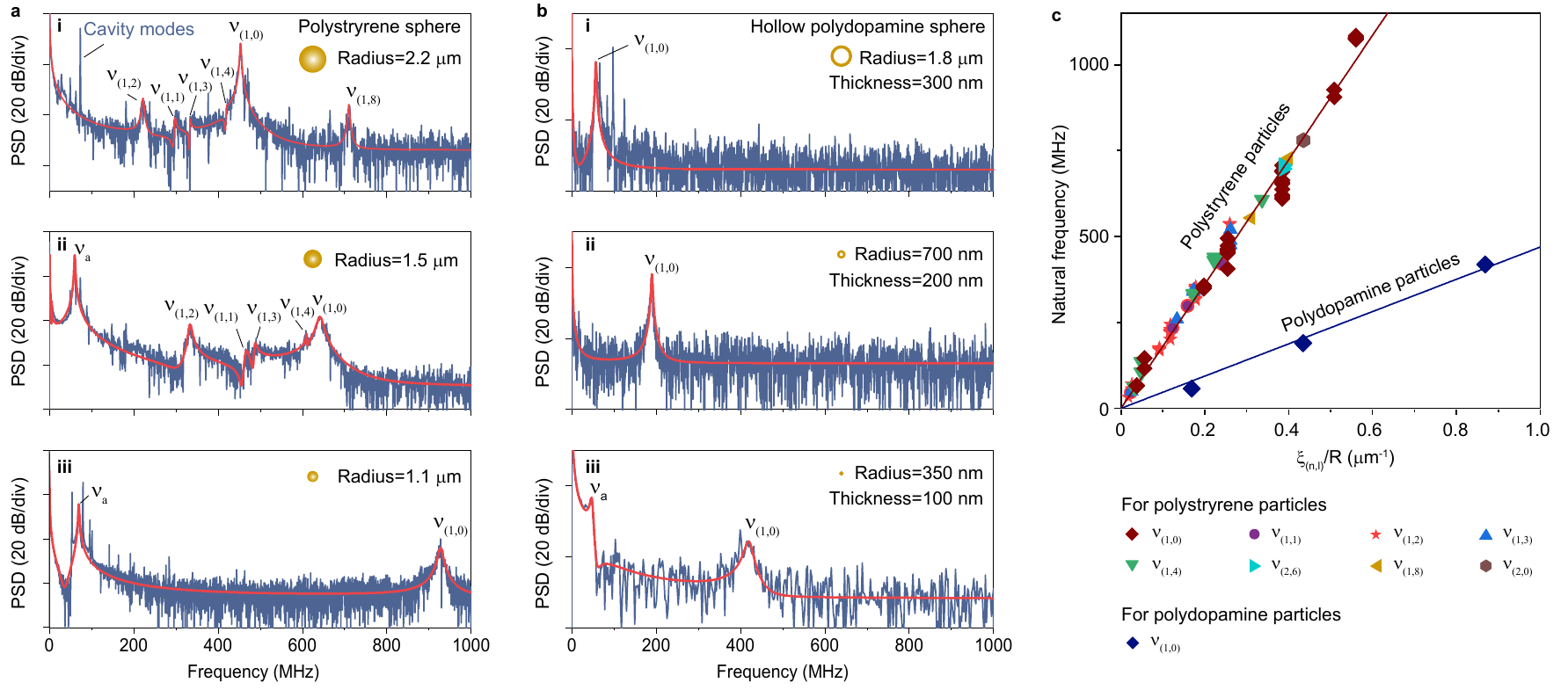}
\caption{\textbf{Mechanical fingerprinting of mesoscopic particles with different sizes and internal structures. a-b}, Vibrational spectra (navy) of single solid polystyrene spheres (\textbf{a}) and hollow polydopamine spheres (\textbf{b}). The red curves are fittings of the natural vibrational modes. Vibrations of the microresonator are also indicated. Insets: the geometry of the particles. \textbf{c}, Statistics of the natural frequencies of polystyrene and polydopamine spheres as a function of $\xi_{(n,l)}/R$. 
}
\label{Fig3}
\end{figure*}

\begin{figure*}[htb]
\centering
\includegraphics[width=13.5cm]{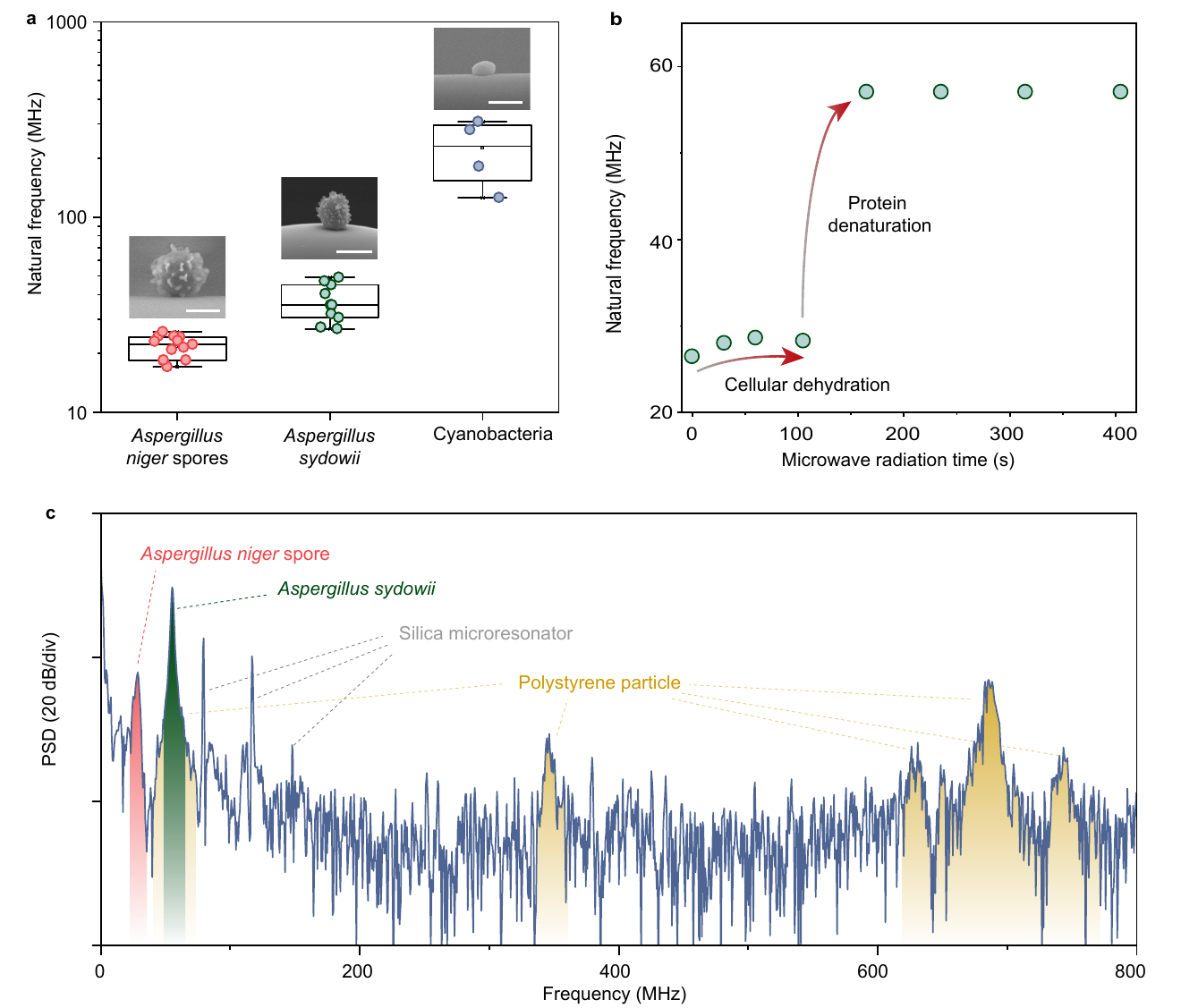}
\caption{\textbf{Biomechanical fingerprinting of microbial cells. a}, 
Statistics of the natural frequencies of the ${(1,2)}$ modes of single {\it Aspergillus niger} spores, living {\it Aspergillus Sydowii} and living cyanobacteria. Boxplots indicate confidence ranges of experimental data (circles). Inset: SEM images of the microorganisms deposited on the microresonator. Scale bars: 2 \textmu m.
\textbf{b}, The natural frequencies of the ${(1,2)}$ mode of a single 
{\it Aspergillus sydowii} as a function of the time exposed to 600 watts microwaves. 
{\bf c}, Vibrational spectra of mixed particles. 
The vibrational modes of the {\it Aspergillus niger} spore, {\it Aspergillus sydowii}, and 1.4-\textmu m-radius polystyrene sphere are indicated by red, olive and yellow respectively. 
}
\label{Fig4}
\end{figure*}

It is first discovered by Pythagoras that the vibrations of strings are drastically enhanced at certain frequencies, which forms the basis of our tone system. 
Such natural vibrations ubiquitously exist in objects regardless of their size scales and have widespread impacts on modern science and industry (Fig.~\ref{fig1}a). 
For example, the stellar oscillations at  millihertz rate have facilitated the determination of interior properties in stars \cite{Chaplin13Annual}.
The crystal oscillators at kilohertz rate have been deployed as time standards in consumer electronics \cite{vittoz2010low}.
In the microscopic world, the molecular vibrations at terahertz rate have become the most common fingerprints for the identification of chemicals and structural analysis of large biomolecules \cite{diem2015modern, coutaz2018principles}.

Recently, natural vibrations of particles at the mesoscopic scale have received growing interest, since this category includes a wide range of functional particles \cite{ lecraw1961extremely, keshtgar2017magnetomechanical,  zhang2016cavity, tayebi2020massively}, as well as most biological cells and viruses \cite{ackerman1951resonances, zinin2005mechanical, dykeman2008low, zinin2009deformation, tsen2012prospects}.
Extending the concept of vibrational spectroscopy to the mesoscopic world enables interrogating particle qualities with regard to their structures and material properties in a non-destructive way \cite{pelton2009damping, crut2014optical, zhou2020single}.
In particular, inferred from the vibrational spectra are important biomechanical properties of cells, which are closely related to their species and living states  \cite{guck2001optical, rojas2018outer, krieg2019atomic}.
However, the acquisition of vibrational spectra of mesoscopic particles is hitherto challenging with existing spectroscopy technologies.  
These particles with sizes ranging from 100 nm to 100 \textmu m are expected to vibrate faintly at megahertz to gigahertz rates \cite{1882lamb, zinin2005mechanical, galstyan2015note}.
Indeed, the natural vibrations at this frequency regime could not be resolved by current Raman and Brillouin spectroscopy due to strong background resulting from Rayleigh-wing scattering in the optical path \cite{wheaton2015probing,  kuok2003brillouin, scarcelli2008confocal}.
Although piezoelectric techniques are widely exploited in macroscopic systems, their performances degrade significantly at frequencies beyond a few megahertz \cite{leisure1997resonant, wissmeyer2018looking}.
In addition, the recent realization of strong coupling between a bacterium and an optomechanical oscillator has limited bandwidth that is inappropriate for practical spectroscopy \cite{gil2020optomechanical}.

Here we demonstrate the real-time measurement of natural vibrations of single mesoscopic particles using an optical microresonator,
extending the reach of vibrational
spectroscopy to a new spectral window.
With the formation of whispering galleries through total-internal-reflections, the microresonator offers high-$Q$ optical resonances that respond to minute changes of the environment \cite{chen2017exceptional, lai2019observation,  yu2021whispering, westerveld2021sensitive, meng2022dissipative}. 
We exploit the photoacoustic effect\cite{rosencwaigphotoacoustics} to stimulate the broadband vibrational modes of particles, which are acoustically coupled to the high-$Q$ optical resonances of the microresonator for real-time detection. 
This mechanism enables extracting the vibrational modes tightly confined inside mesoscopic particles that have not been resolved using standard photoacoustic technologies \cite{wang2016practical, strohm2015single, tan2011multimodal, strohm2013probing,strohm2015classification}. 
Using this mechanism, single-particle vibrational spectroscopy is realized with SNRs over 50 dB and a detection bandwidth beyond 1 GHz. 
Remarkably, we unveil the vibrational spectra of single microbial cells, which are found to be highly related to their species and living states. 
Our implementation allows vibrational spectroscopy of a wide range of mesoscopic particles, which could revolutionarily advance our understanding of the mesoscopic world with unprecedented precision.

The working principle of microresonator-based vibrational spectroscopy is illustrated in Fig. \ref{fig1}b, in which the underlying idea is to transfer the particle vibrations to the frequency modulation of optical resonances. As shown in the inset ii of Fig. \ref{fig1}b, the particle is deposited onto a whispering-gallery-mode microresonator and then irradiated by a pulsed laser (200 ps, 532 nm). Upon absorption of a short laser pulse, the acoustic transient pressure is generated inside the particle due to thermoelastic expansion.
The transient pressure excites vibrational modes of the particle over a broad bandwidth,
which are coupled to the optical resonance by stimulating acoustic waves in the microresonator. When the acoustic waves propagate to approach the optical mode, they alter the refractive index and the boundary of the microresonator, and thereby modulate the frequency of the optical resonance.
A continuous-wave probe laser coupled to the microresonator by a microfiber is slightly detuned from the optical resonance to enable real-time readout. 
The vibrational spectra are derived from the power spectral density of the transmitted laser.

Standard spherical polystyrene particles are first tested to benchmark the detection performance. A single particle is deposited on the silica microspherical resonator with 
the radius of 30 \textmu m.  
By setting the energy density of the incident pulse to $2$ pJ \textmu m$^{-2}$, oscillations are observed in the transmission of probe laser (Fig.~\ref{Fig2}a), which gradually damp in a few microseconds. The vibrational spectrum is derived by Fourier transform of 
the first 3.2 \textmu s of the trace, and multiple peaks are revealed with frequencies up to 800 MHz (Fig.~\ref{Fig2}b). 
Due to the efficient acousto-optic coupling and the high-$Q$ optical resonance ($\sim 10^6$), the signal-to-noise-ratio (SNR) is observed up to 50 dB. 
The corresponding vibrational modes $\nu_{\text{(n,l)}}$ are identified by matching spectral features to theoretical predictions, with ($n$,$l$) being the radial and angular mode numbers  \cite{1882lamb} 
(see Extended Data Fig.~1, Methods and Supplementary Note I for details of theoretical analysis). Note that, compared with free particles, the contact with the microresonator causes  slight frequency shifts of natural vibrations ($<5$\%, see Supplementary Fig.~1), and introduces a new vibrational mode ($\nu_{\text{a}}$). Besides, the damping rates of each vibrational mode can be also obtained. By using short-time Fourier transform, 
the evolution of the intensity of the peaks is extracted and plotted in Fig.~\ref{Fig2}c, which decays exponentially over time. The mechanical quality factors of these vibrational modes are within 30 to 200.

We further investigate optical sensitivity of the microresonator. 
For vibrational modes with frequency $\nu$, the sensitivity should follow the theoretical trend (see Supplementary Note II)
\begin{equation}
S=\alpha  Q_{\text{o}} \sqrt{\frac{1+4( \nu Q_{\text{o}} /f_{\text{o}})^2}{1+4(\nu Q_{\text{o}} /f_{\text{o}})^4}},  
\end{equation}
where $Q_{\text{o}}$ and $f_{\text{o}}$ represent the quality factor and resonant frequency of the optical mode, respectively. 
The coefficient $\alpha$ is determined by the geometry and material of the microresonator. 
As shown in Fig. \ref{Fig2}d, the sensitivity approaches its maximum value of $\alpha f_{\text{o}}/\nu$ for vibrations at frequencies beyond the microresonator linewidth ($\nu> f_{\text{o}}/Q_{\text{o}}$).
Therefore, devices with higher $Q_{\text{o}}$ provide optimal sensitivity over a broader spectral window.
This theoretical formula is tested experimentally by tuning the $Q_{\text{o}}$ of the microresonator using an extra scatterer. 
The vibrational spectra of a 1.4-\textmu m-radius polystyrene particle are measured. The extracted peak intensities at three natural frequencies are shown in Fig.~\ref{Fig2}e, from which the optical responses increase with $Q_{\text{o}}$ until approaching their upper limit, in good agreement with the theoretical prediction. 
Besides, ultrasound-mediated coupling makes it possible to detect distant particles without spoiling $Q_{\text{o}}$ of optical resonances. In experiments, we place four particles at different locations on the microresonator and its substrate (Figs.~\ref{Fig2}f-g). 
Particle vibrations are observed in laser transmissions for all four cases (Fig.~\ref{Fig2}f). Note that, the distance between the optical mode and the particles causes a delay between the pulse excitation and its optical readout.
The measured time delay is plotted as a function of the distance, from which the acoustic velocity is derived as 5.96 km s$^{-1}$ in agreement with the longitudinal acoustic velocity in silica (Fig.~\ref{Fig2}g). 

Figure~\ref{Fig3}a shows vibrational spectra of three polystyrene spheres with different sizes.
The breathing modes ($\nu_{(1,0)}$) have the largest cross sections for pulse excitation and appear as the primary peaks in the spectra (see Methods and Extended Data Fig.~1 for theoretical analysis). Their natural frequencies are found to be inversely proportional to the radius as predicted theoretically\cite{1882lamb} that $\nu_{\text{(n,l)}}=\sqrt{E/\rho}\times\xi_{\text{(n,l)}}/R$
with $R$ the radius of the particle, $E$ the Young's modulus, $\rho$ the material density, and $\xi_{\text{(n,l)}}$ the dimensionless factor (see Supplementary Fig. 1). 
The same tendency is shown in Fig.~\ref{Fig3}b for polydopamine spheres with hollow internal structures. 
It is worth noting that the vibrational modes of microresonator itself can also be excited by the particle and radiation pressure \cite{Ma07, gil2020optomechanical}, which are manifested as narrow peaks (Extended data Fig.~2). These spectral information further provide a measure of the mechanical property of the particles. Figure~\ref{Fig3}c presents the measured natural frequencies of two types of particles as a function of $\xi_{\text{(n,l)}}/R$. In light of the material densities of polystyrene ($1.05$ g~cm$^{-3}$)  and polydopamine ($1.52$ g~cm$^{-3}$), their Young's moduli are estimated from fitted slopes to be $3.5$ GPa and $0.3$ GPa respectively.

Now we explore the feasibility for biomechanical fingerprinting of microbial cells. Three types of microorganisms, the living cyanobacteria ({\it Synechocystis} sp.~PCC6803), living {\it Aspergillus sydowii} and {\it Aspergillus niger} spores, are deposited onto the microresonator's surface (see Methods). The pulse energy density is set to be less than 5 pJ \textmu m$^{-2}$ to avoid photodamage.  
Their temporal and spectral characterizations could be obtained routinely, showing an SNR exceeding 20 dB (see Extended Data Fig. 3). 
Figure~\ref{Fig4}a summarizes the natural frequencies of the quadrupole mode $\nu_{\text{(1,2)}}$, with mean values of $21.6\pm2.9$ MHz, $36.5\pm8.4$ MHz, and $223.0\pm70.0$ MHz for {\it Aspergillus niger} spores, {\it Aspergillus sydowii} and the cyanobacteria. The bunching of natural frequencies for microorganisms belonging to the same species demonstrates their potential as unique fingerprints, a result of the highly defined and stable morphology of certain biological species. The slight variation in natural frequencies may provide information of the living states of the microorganisms. It is tested by measuring the vibrational spectra of a (living) {\it Aspergillus sydowii} under certain time of microwave radiation (2.450 GHz, 600 W). 
Figure \ref{Fig4}b presents the evolution of natural frequency  $\nu_{\text{(1,2)}}$ over the radiation time, which firstly shows a gradual increase by 8\% from 26.5 MHz due to the drying of free water, and then a rapid transition to 57 MHz resulting from protein denaturation\cite{gil2020optomechanical, woo2000differential, pillet2014uncovering} (see Supplementary Note III and Supplementary Fig.~2). 
Besides, the mixed particles are distinguished by this technology. An {\it Aspergillus niger} spore, an {\it Aspergillus sydowii} and a 1.4-\textmu m-radius polystryrene sphere are deposited onto the microresonator in sequence. Their vibrations can be excited solely or simultaneously by adjusting the spot size of the pulsed laser (see Supplementary Fig.~3). Characteristic peaks corresponding to different particles are clearly observed and are overlaid in the spectra (Fig.~\ref{Fig4}c). The high SNRs of the peaks allow for distinguishing the vibrational features of these particles. 

To conclude, we have demonstrated the real-time measurement of natural vibrations of single mesoscopic particles using an optical microresonator,
extending the reach of vibrational
spectroscopy to a new spectral window. The detection capability of this technology was testified by particles with different constituents, sizes, and internal structures.
The detection bandwidth exceeding 1 GHz meets the requirements for a wide range of mesoscopic particles, offering invaluable opportunities to investigate biomechanics and nanomechanics at the single-particle level\cite{yu2016cavity, sansa2020optomechanical, sbarra2022multimode}. We applied this technology for the biomechanical fingerprinting of microbial cells with different species and living states.
The acquisition speed of this technology is only limited by the damping time of the natural vibrations, meeting the requirement for rapid dynamic analysis of mesoscopic particles, such as the chemical synthesis of various particles and the growth of biological cells.
While proof-of-concept experiments are performed in air, the ultrasonic detection scheme is fundamentally compatible with microfluidic systems\cite{Han:16, fan2014potential, galstyan2015note}, holding great potential in single-particle flow cytometers.
Besides, as the natural frequencies of biological cells could be clearly resolved, this technology provides crucial guidelines for selectively destroying microorganisms or cancer cells through resonant ultrasound and microwaves without damaging healthy cells \cite{mittelstein2020selective, wierzbicki2021effect, sadraeian2022viral}.


%

\vspace{20pt}
\noindent \textbf{\LARGE{Methods}}

\medskip
{\bf\noindent Theoretical model}\\
The absorption of a short laser pulse generates time-variant acoustic pressure by thermoelastic expansion inside objects \cite{xu2006photoacoustic}, stimulating their natural vibrations. To analyze this mechanism, a theoretical model is build up in Supplementary Note I. When the duration time $\tau$ of the laser pulse is so short that thermal diffusion can be neglected,  
the vibration amplitude of the eigen-mode is derived as
\begin{equation}
    A_{j}=\left|d_{j}\right| e^{-\frac{1}{2} \omega_{j}^{2} \tau^{2}-\gamma_j t},
    \label{amplitude after excitation}
\end{equation}
where $\omega_j$ and $\gamma_j$ denote the angular frequency and damping rate of the $j$-th vibrational mode, respectively;
$d_{j}=\int\boldsymbol{u}_{j}(\boldsymbol{r}) \cdot \boldsymbol{u}_{\text{e}}(\boldsymbol{r})  \mathrm{d}^{3} \boldsymbol{r}$ represents the spatial overlap of normalized mode displacement field  $\boldsymbol{u}_{j}(\boldsymbol{r})$ with that imposed by the absorption-induced thermal expansion $\boldsymbol{u}_{\text{e}}(\boldsymbol{r})$. 

The time-domain vibrations of a homogeneous spherical particle triggered out by a laser pulse is theoretically analyzed (Extended Data Fig.~1). The validity of the analytical model is confirmed by three-dimensional finite-element simulation, in which the thermal diffusion and thermoelastic damping existed in the real scenario are considered. 
Generally, within the excitation bandwidth $1/\tau$, the breathing mode (1,0) of a homogeneous spherical particle has the largest efficiency, while other spheroid modes can be also stimulated by the nonuniform thermal distribution induced by the light focusing effect inside the particle \cite{li2005optical}.

\vspace{6pt}
\noindent
\textbf{Experimental setup}\\
A pulsed laser (Standa STA-01-MOPA-SH-3, wavelength 532 nm, pulse width 200 ps, repetition rate 1 Hz-1 kHz) is passed through a variable neutral-density filter to adjust its power, a silver mirror controlled via a three-dimensional translation stage to scan its position, and focused onto the target particles with a microscope objective (20$\times$, NA=0.4). The probe light from a tunable continuous-wave diode laser (Toptica, 1550 nm) is coupled to the microresonator via a contacting tapered fiber, and its transmission is received by a 45-GHz-bandwidth photodetector. The detected electrical signals are amplified (SHF 806E, 26 dB at 40 kHz-38 GHz) and recorded by an oscilloscope. To improve the SNRs, 64 traces are averaged in a single measurement. A time-windowed portion of the oscillatory signals is chosen for Fourier transformation according to the desired SNRs and spectral resolution (3.2 \textmu s for most cases, 0.4 \textmu s for the weak signals of the 350-nm-radius hollow polydopamine sphere in Fig. 3b-iii and the living cyanobacterium in Extended Data Fig. 3b-iii).

\vspace{6pt}
\noindent
\textbf{Sample preparation}\\
The dye-doped polystyrene spheres (mean radii: 2.8 \textmu m, 2.2 \textmu m, 1.5 \textmu m and 1 \textmu m; Xi'an ruixi Biological Technology Co., Ltd) and hollow black polydopamine spheres (mean radii, 1.8 \textmu m, 700 nm and 350 nm with the thicknesses of about 300 nm, 200 nm and 100 nm respectively; Xi'an Confluore Biological Technology Co., Ltd.) are purchased from commercial suppliers. The suspensions of these particles are treated by ultrasonication, and then transferred onto the microresonator's surface via a fiber microtip.

The living cynobacteria ({\it Synechocystis} sp.~PCC6803), {\it Aspergillus niger} spores (ATCC 16404) and living {\it Aspergillus sydowii} (ON408995.1) are cultured on a growth medium solidified with agar. To transfer them onto the microresonator, the microorganisms are spread out on the agar medium, and the spherical microresonantor on a standard silica fiber functions as a probe to adsorb the monodisperse microbial particles. Then, the measurements are performed in atmospheric environment shortly.

\vspace{6pt}
\noindent \textbf{Acknowledgment}\\
We thank Dr. Xiao-Chong Yu, Prof. Changhui Li and Prof. Jianyong Huang for the helpful discussions. S.-J.T., J.S. and Y.F.X. thank Prof. Jindong Zhao, and Yaqing Ji at Peking University for help in bacteria experiments.
This project is supported by 
the National Natural Science Foundation of China (Grant Nos. 11825402, 12041602, 11654003 and 62105006), and the High-performance Computing Platform of Peking University. S.-J.T. is supported by the China Postdoctoral
Science Foundation (Grant Nos. 2021T140023 and 2020M680187).
\noindent

\renewcommand{\figurename}{\textbf{Extended Data Fig.}}
\begin{figure*}[hbtp]
\centering
\setcounter{figure}{0}
\includegraphics[width=\linewidth]{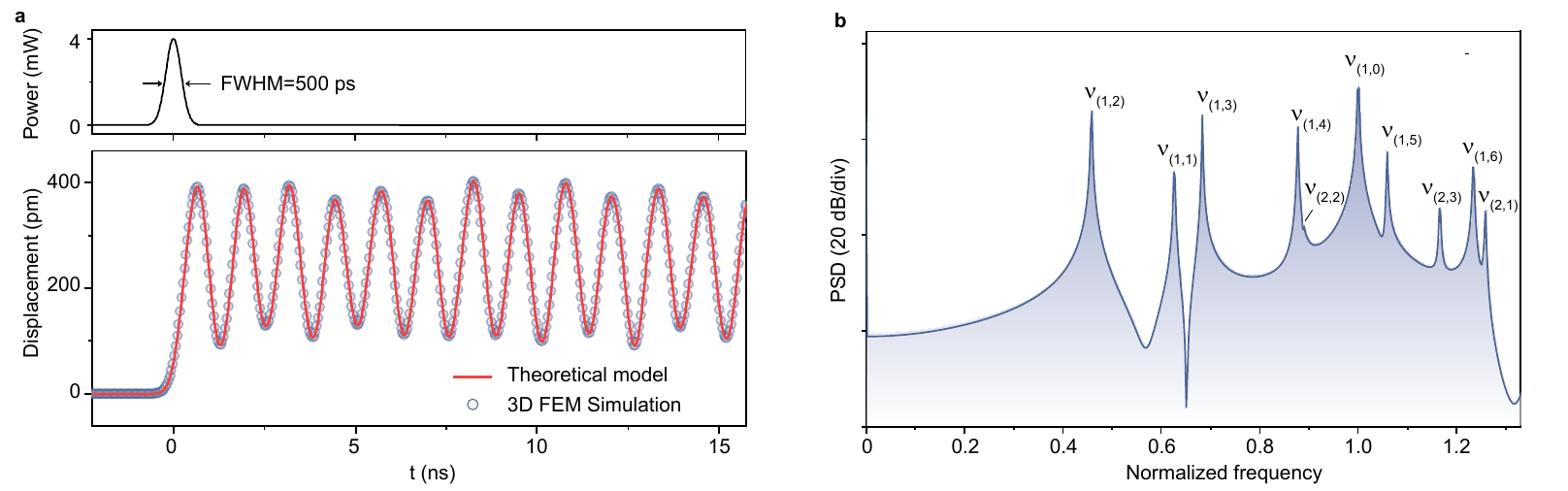}
\caption{\textbf{Photoacoustic excitation  of particle's vibrations. a}, Light absorption of a short laser pulse excites mechanical resonances of the particle. Top panel: the laser pulse; Bottom panel: the particle's surface displacement (Curve: theoretical model; Circles: three-dimensional finite-element method).
\textbf{b}, Typical vibrational spectrum of the spherical particle irradiated by a pulsed laser. The frequency is normalized to the natural frequency of the (1,0) mode.}
\label{pic:ED1}
\end{figure*}

\begin{figure*}[hbtp]
\centering
\setcounter{figure}{1}
\includegraphics[width=14cm]{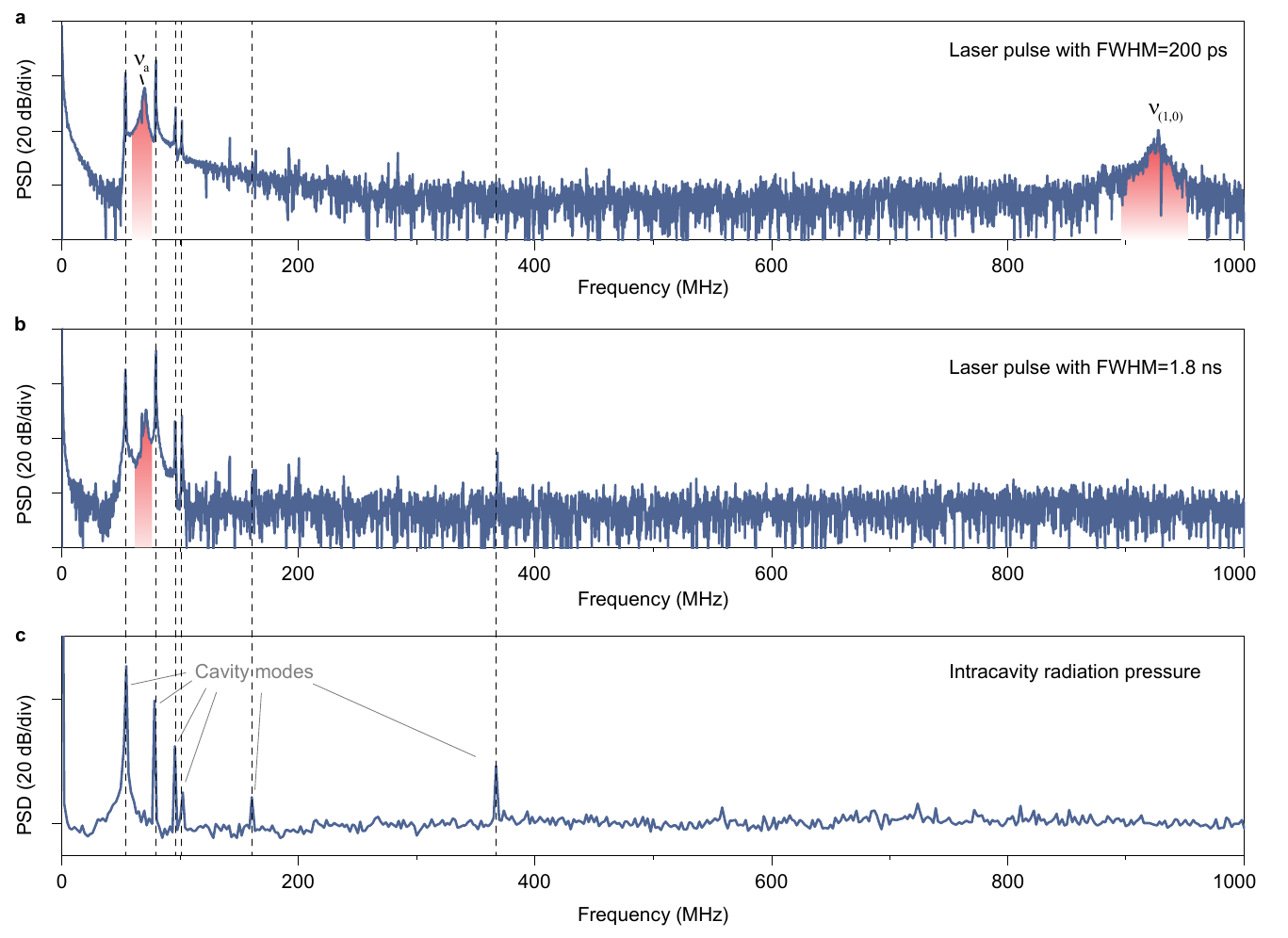}
\caption{\textbf{Identifications of vibrational modes. a-b}, Vibrational spectra obtained when a 1.1-\textmu m-radius polystyrene sphere is deposited on the microresonator's surface and irradiated by laser pulses with the pulse duration of 200 ps (a), and 1.8 ns (b), respectively. Vibrational modes of the particle are indicated with the red shadows. 
\textbf{c}, Vibrational spectra obtained with the intracavity radiation pressure\cite{Ma07} to identify the vibrational modes of the silica microsphere (dashed lines labelled).
}
\end{figure*}

\begin{figure*}[hbtp]
\centering
\setcounter{figure}{2}
\includegraphics[width=18cm]{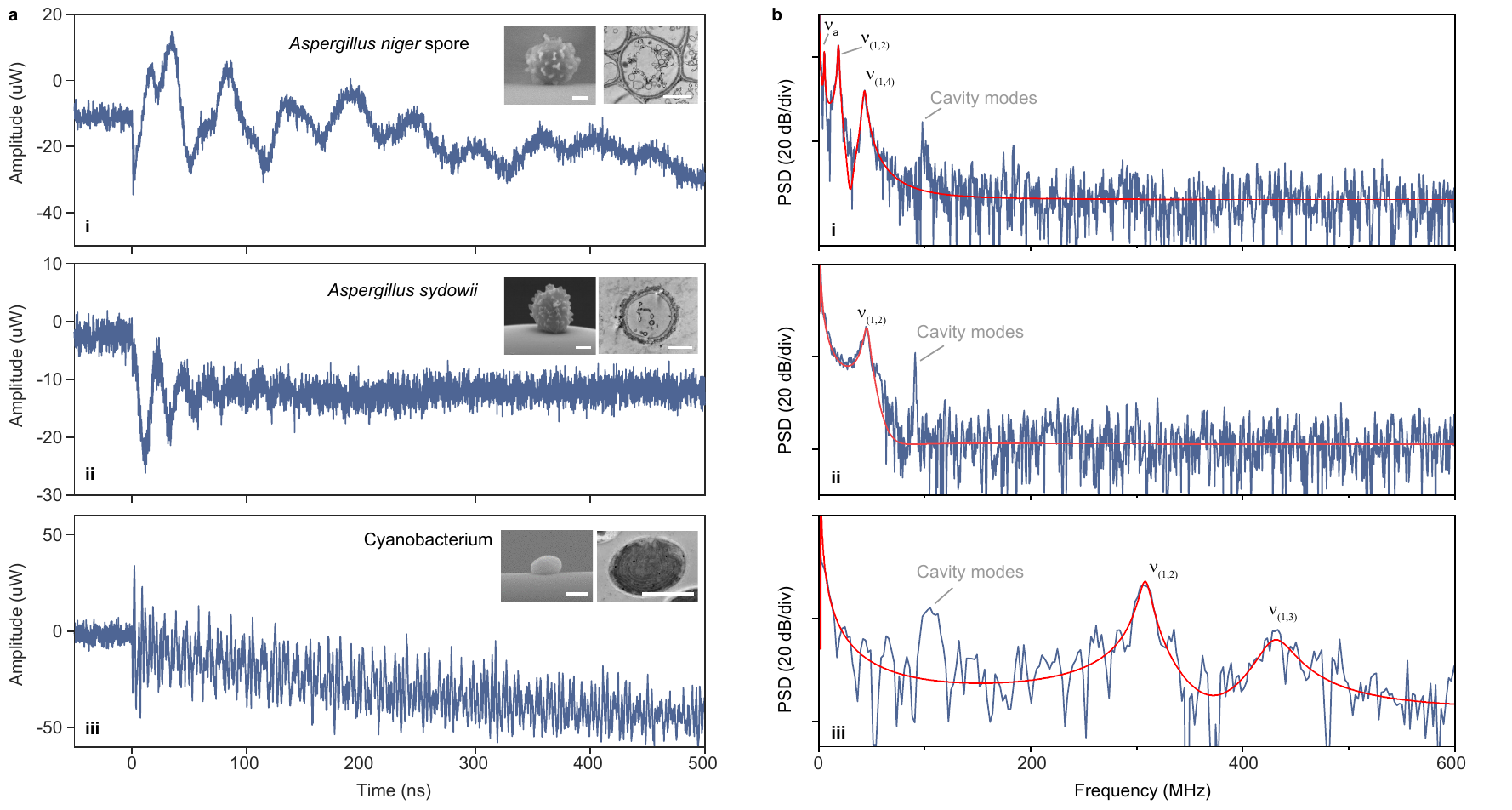}
\caption{\textbf{Temporal and spectral characterization of natural vibrations of single microbial cells. } Laser transmission (\textbf{a}) and corresponding vibrational spectra (\textbf{b}) of a single {\it Aspergillus niger} spore (i), {\it Aspergillus Sydowii} (ii) and cyanobacterium (iii). The navy and red curves are experimental data and fittings, respectively. Vibrations of the microresonator are also indicated. The mode indices of these vibrational modes are identified by matching the spectral features to numerical calculations. Insets: SEM and transmission-electron-microscope (TEM) images of microbial cells. Scale bars: 1 \textmu m. 
}
\end{figure*}

\clearpage

\setcounter{equation}{0}
\renewcommand{\theequation}{S\arabic{equation}}
\begin{widetext}
\noindent \textbf{\LARGE{Supplementary Information}}\\

{\bf\noindent I. Excitation mechanism of natural vibrations}\\
When an object is irradiated by a short laser pulse, the light scattering determines the optical field distribution inside the object. As the time scale of electromagnetic (EM) field to build up is very fast ($\ll 1$ ps for a microscale particle), 
the optical field distribution inside the particle can be described by $I(\boldsymbol{r})e^{-t^2/2\tau^2}$ with $\tau$ being the pulse width. 
The heat flux inside the particle induced by electromagnetic (EM) absorption is thus determined as $Q(\boldsymbol{r}, t)=q(\boldsymbol{r})e^{-t^2/2\tau^2}$, with $q(\boldsymbol{r})=\alpha I(\boldsymbol{r})$ being the absorbed heat distribution at the peak power point of laser pulse and $\alpha$ being the absorption coefficient of the object.

To stimulate efficiently natural vibrations of particles, the duration time $\tau$ of the laser pulse should be short enough that thermal diffusion can be neglected ($\tau\ll\tau_{\text{th}}$). The time scale for heat dissipation of absorbed EM energy by thermal conduction is estimated as $\tau_{\text{th}}= \rho C_p L^{2} /4 \kappa$ with $\rho$  material density(suppose the particle is made up of homogeneous material in the model), $C_p$ the heat capacity (per volumn) and $\kappa$ the thermal conduction coefficient and $L$ the characteristic linear dimension of the particle \cite{xu2006photoacoustic} ($\tau_{\text{th}}>100$ ns for a microscale organic particle). 
Under this condition, the temperature rise $T(\boldsymbol{r}, t)$ induced by the EM absorption can be derived as
\begin{equation}
T(\boldsymbol{r}, t)=\frac{1}{\rho C_{p}} \int_{-\infty}^{t} Q(\boldsymbol{r}, t^\prime) \mathrm{~d} t^\prime =\frac{1}{\rho C_{p}} q(\boldsymbol{r}) \int_{-\infty}^{t}  e^{-\frac{t^{\prime 2}}{2 \tau^{2}}} \mathrm{~d} t^\prime = T_0(\boldsymbol{r}) (1+\text{erf}(\frac{t}{\sqrt{2}\tau})) /2,
\label{temp}
\end{equation}
where erf is error function, $T_0(\boldsymbol{r})=\frac{\sqrt{2\pi}\tau}{\rho C_{p}} q(\boldsymbol{r})$ denotes the temprature rise after the laser pulse excitation.

The induced thermal stress acts as an additional volume force imposed on the particle, and the motion of the particle can be described by the modified elastic wave equation as \cite{Landau1986}
\begin{equation}
\rho \frac{\partial^{2}}{\partial t^{2}} \boldsymbol{u}=\mathcal{L} \boldsymbol{u}-K \alpha \nabla T,
\label{eom}
\end{equation}
where $\boldsymbol{u}$ is the displacement field,  $\alpha$ is thermal expansion coefficient, $\mathcal{L}=(\lambda+2 \mu) \nabla(\nabla \cdot)-\mu \nabla \times(\nabla \times)$,  $\lambda={E \sigma}/[{(1-2 \sigma)(1+\sigma)}]$ ,  $\mu={E}/[{2(1+\sigma)}]$ and $K=\lambda+2\mu/3$ with Young's modulus $E$ and Poisson's ratio $\sigma$.

Thermal expansion also alters the form of boundary conditions. Consider a particle attached on the rigid substrate, fixed boundary condition is appropriate for the contact part of the particle, while the non-contact part remains free boundary condition. With thermal expansion taken into consideration, the relation between the boundary displacement field $u_i$ and the stress tensor $\sigma_{i k}$ imposing on the boundary satisfy
\begin{equation}
\begin{cases}
\sigma_{i k} n_k =\left[-K \alpha T \delta_{i k}+K u_{l l} \delta_{i k}+2 \mu\left(u_{i k}-\frac{1}{3}  u_{l l} \delta_{i k} \right) \right] n_k =0  &{\text{Free part}}\\
u_k n_k =0 &{\text{Contact part}} \\
\end{cases}
\label{bc}
\end{equation}
Here $n_k$ is the $k$-component of the normal vector, $u_{i k}=\frac{1}{2}\left(\frac{\partial u_{i}}{\partial x_{k}}+\frac{\partial u_{k}}{\partial x_{i}}\right)$ is the strain tensor, and Einstein summation rule is adopted here.

Combining Eq.~\ref{eom} and Eq.~\ref{bc}, along with Eq.~\ref{temp}, the motion of the particle can be solved, which is decomposed mathematically into two parts $\boldsymbol{u}=\boldsymbol{u}_{\text{e}}+\boldsymbol{u}_{\boldsymbol{v}}$:  

\noindent
(1) The former $\boldsymbol{u}_{\text{e}}$ represents the time evolution of the equilibrium position induced by thermal expansion. Temporally, $\boldsymbol{u}_{\text{e}}$ depends linearly on temperature as
\begin{equation}
\boldsymbol{u}_{\text{e}} (\boldsymbol{r},t)= \boldsymbol{u}_{\text{e}} (\boldsymbol{r})(1+\text{erf}(\frac{t}{\sqrt{2}\tau})) /2,
\end{equation}
where $\boldsymbol{u}_{\text{e}} (\boldsymbol{r})$, the equilibrium position after the pulse excitation, can be derived by solving the stable state equation
\begin{equation}
\mathcal{L}\boldsymbol{u}_{\text{e}}(\boldsymbol{r}) -K \alpha \nabla T_0(\boldsymbol{r})=0
\label{eq1}
\end{equation}
with the boundary condition 
\begin{equation}
\begin{cases}
\sigma_{e, i k} n_k=\left[ -K \alpha T_0(\boldsymbol{r}) \delta_{i k}+K u_{e, l l} \delta_{i k}+\left.2 \mu\left(u_{e, i k}-\frac{1}{3} \delta_{i k} u_{e, l l}\right)\right. \right] n_k=0 &\text{Free part}\\
u_{e,k}n_k=0 &\text{Contact part}\\
\end{cases}
\label{eq2}
\end{equation}
Note that the solution to Eq.\ref{eq1} and Eq.\ref{eq2} is the steady-state solution to thermal expansion under given temperature field $T_0(\boldsymbol{r})$.

\noindent
(2) The latter $\boldsymbol{u}_{\boldsymbol{v}}$ depicts the vibration around the equilibrium position. It satisfies the equation of motion
\begin{equation}
\begin{aligned}
\rho \frac{\partial^{2}}{\partial t^{2}} \boldsymbol{u}_{\boldsymbol{v}} &=\mathcal{L} \boldsymbol{u}_{\boldsymbol{v}}-\rho \frac{\partial^{2}}{\partial t^{2}} \boldsymbol{u}_{\boldsymbol{e}}(\boldsymbol{r},t) \\
&=\mathcal{L} \boldsymbol{u}_{\boldsymbol{v}}+\rho \boldsymbol{u}_{\boldsymbol{e}}(\boldsymbol{r}) \frac{t}{\sqrt{2\pi}\tau^3}e^{-t^2/{2\tau^2}},
\end{aligned}
\end{equation}
and the homogeneous boundary condition
\begin{equation}
\begin{cases}
\sigma_{v, i k} n_k =\left[ K u_{v, l l} \delta_{i k}+\left.2 \mu\left(u_{v, i k}-\frac{1}{3} \delta_{i k} u_{v, l l}\right)\right. \right]n_k=0 &\text{Free part}\\
u_{v,k}n_k=0 &\text{Contact part}\\
\end{cases}
\end{equation}

The operator $\mathcal{L}$ is an Hermitian operator when the boundary condition is homogeneous \cite{sudhir2017quantum}. Therefore, $\boldsymbol{u}_{\boldsymbol{v}}$ can be decomposed into the motion of eigen-modes

\begin{equation}
    \boldsymbol{u}_{\boldsymbol{v}}(\boldsymbol{r}, t)= \sum_j {a_{j}(t) \boldsymbol{u}_{j}(\boldsymbol{r})}
\end{equation}
with $a_{j}(t)$ and  $\boldsymbol{u}_{j}(\boldsymbol{r})$  being the amplitude and the normalized displacement field of the $j$-th vibrational eigen-mode of the particle, respectively. The equation of motion for each mode is derived simply as
\begin{equation}
\frac{\mathrm{d}^{2} a_{j}}{\mathrm{d} t^{2}}+\omega_{j}^{2} a_{j}=d_{j}\frac{t}{\sqrt{2\pi}\tau^3} e^{-t^2/{2\tau^2}},
\end{equation}
where $d_{j}= \int\boldsymbol{u}_{j}(\boldsymbol{r}) \cdot \boldsymbol{u}_\text{e}(\boldsymbol{r}) \mathrm{d}^{3} \boldsymbol{r}$ characterizes the spatial overlap of normalized mode displacement field  $\boldsymbol{u}_{j}(\boldsymbol{r})$ with that imposed by the absorption-induced thermal expansion $\boldsymbol{u}_{\text{e}}(\boldsymbol{r})$, and $\omega_j$ is the angular frequency of the $j$-th vibrational eigen-mode of the particle.

Therefore, the vibration amplitude for the $j$-th mode after the pulse excitation is derived as
\begin{equation}
A_{j}=\left|d_{j}\right| e^{-\frac{1}{2} \omega_{j}^{2} \tau^{2}}.
\end{equation}
Furthermore, consider the energy dissipation rate  of mechanical mode $\gamma_j$ that is typically slow enough to be neglected in the transient excitation process, and the vibration amplitude after the excitation can be written as
\begin{equation}
A_{j}(t)=\left|d_{j}\right| e^{-\frac{1}{2} \omega_{j}^{2} \tau^{2}-\frac{1}{2} \gamma_j t}.
\end{equation}

\clearpage
{\bf\noindent II. High-$Q$-microresonator-enhanced sensitivity}

\noindent In the microresonator-based vibrational spectrum measurement, 
the vibrational modes of the particle with oscillation amplitude $A$ and frequency $\omega=2\pi\nu$ are acoustically coupled to the microresonator via mechanical contact, which stimulates acoustic waves to alter the refractive index and the boundary of the microresonator; Such acoustic modulation can be equivalent to the modulation on the effective refractive index $n_{\text{eff}}$ of the optical mode, and leads to a perturbation on its resonant frequency $f_{\text{o}}$, which is read out in the transmission $T$ of a probe light slightly detuned from the optical resonance with $\Delta$. 
Therefore, the detection sensitivity of the microresonator to particle vibrations is written as $S(\omega)= | \chi_{\mathrm{m}} | S_{\mathrm{o}}$ with the mechanical susceptibility $\chi_{\mathrm{m}}=\frac{\delta n_{\text{eff}}}{A}$ and the optical  sensitivity $S_{\mathrm{o}}=| \frac{ \delta T }{ \delta \Delta }\frac{\delta (2 \pi f_{\text{o}})}{\delta n_{\text{eff}}} | $. Here, the mechanical susceptibility $\chi_{\text{m}}$ is determined mainly by the acoustic coupling between the particle and the microresonator, as well as the distance between the particle and the optical mode.

In the following, we analyze the optical sensitivity of the microresonator. Theoretically, vibration-induced acoustic waves impose a temporal perturbation on the frequency detuning $\delta \Delta(t)$ between the probe light and the optical mode, which can be written as
\begin{equation}
\Delta(t)=\Delta_{0}+\delta \Delta(t),
\label{detunig}
\end{equation}
with its Fourier transform being
\begin{equation}
\delta {\Delta}(\omega)=\int \delta \Delta(t) e^{i\omega t} \mathrm{d} t.
\end{equation}

The equation of motion for the microresonator-microfiber coupled system \cite{gorodetsky1999optical} can be written as 
\begin{equation}
\dot{a}=-\frac{\kappa}{2} a+i \Delta a+\sqrt{\kappa_{ex}} a_{\text{in}},
\label{input-ouput}
\end{equation}
where $a$ is the optical field inside the cavity, $a_{\text{in}}$ is the input optical field,
$\kappa=\kappa_0+\kappa_{\text{ex}}$ with $\kappa_0$ and $\kappa_{\text{ex}}$ being the intrinsic loss and coupling dissipation of the cavity mode, respectively. 
The stable solution of the cavity optical field can be acquired as
\begin{equation}
a_{0}=\frac{\sqrt{\kappa_{\text{ex}}} a_{\text{in}}}{\frac{\kappa}{2}-i \Delta_{0}}.
\label{stable}
\end{equation}

Suppose acoustic modulation results in a perturbation $\delta a(t)$ from the stable cavity optical field $a_{0}$ as
\begin{equation}
a(t)=a_{0}+\delta a(t),
\label{cavity field}
\end{equation}
with its Fourier transform
\begin{equation}
\delta {a}(\omega)=\int \delta a(t) e^{i\omega t} d t.
\end{equation}
By substituting Eqs.~\ref{detunig}, ~\ref{cavity field} and \ref{stable} into  Eq.~\ref{input-ouput}, the input-output equation becomes
\begin{equation}
{\delta \dot{a}(t)}=-\frac{\kappa}{2} \delta a(t)+i \delta \Delta(t) a_{0}+i \Delta_{0} \delta a(t),
\end{equation}
which is equivalent to 
\begin{equation}
\delta {a}(\omega)=\frac{i a_{0} \delta {\Delta}(\omega)}{\frac{\kappa}{2}-i\left(\omega+\Delta_{0}\right)}.
\end{equation}

According to the input-output relation 
$a_{ \text{out}}= a_{\text{in}}-\sqrt{\kappa_{\text{ex}}} a 
=a_{\text{in}}-\sqrt{\kappa_{\text{ex}}} a_{0}-\sqrt{\kappa_{\text{ex}}} \delta a  \left(t\right)$, 
the transmission of the microresonator-microfiber coupled system is obtained as
\begin{equation}
\begin{aligned}
T(t)=\frac{\left|a_{ {\text{out} }}(t)\right|^{2}}{\left|a_\text{in}\right|^{2}}=&\left| 1-\frac{\kappa_\text{ex}}{\frac{\kappa}{2}-i \Delta_{0}}\right|^{2}\\
&-\left(1-\frac{\kappa_\text{ex}}{\frac{\kappa}{2}-i \Delta_{0}}\right)^{*}\left(\frac{\kappa_\text{ex}}{\frac{\kappa}{2}-i \Delta_0} \int \frac{1}{2 \pi} \frac{i}{\frac{\kappa}{2}-i\left(\omega+\Delta_{0}\right)} \delta \Delta(\omega) e^{-i \omega t}\right)\\
&-\left(1-\frac{\kappa_\text{ex}}{\frac{\kappa}{2}-i \Delta_{0}}\right)^{}\left(\frac{\kappa_\text{ex}}{\frac{\kappa}{2}-i \Delta_0} \int \frac{1}{2 \pi} \frac{i}{\frac{\kappa}{2}-i\left(\omega+\Delta_{0}\right)} \delta \Delta(\omega) e^{-i \omega t}\right)^*.\\
\end{aligned}
\end{equation}
Here $\int e^{i(\omega -\omega^\prime)t}/2\pi= \delta(\omega-\omega^\prime)$; Since $\ \delta \Delta(t)\ $ is real, we have $\delta \Delta(-\omega) ^* =\delta \Delta(\omega)$. 

Then, the Fourier transform of the transmission can be obtained, whose amplitude $| \delta T(\omega)| $ is (the DC term corresponding to the unperturbed response is ignored)
\begin{equation}
\begin{aligned}
| \delta T(\omega) | 
=\frac{2 \Delta_{0} \kappa_\text{ex} \sqrt{(\kappa-\kappa_\text{ex})^{2}+\omega^{2}}}{\left(\frac{\kappa^{2}}{4}+\Delta_{0}^{2}\right) \sqrt{\left(\frac{\kappa^{2}}{4}+\Delta_{0}^{2}-\omega^{2}\right)^{2}+\omega^{2} \kappa^{2}}} |\delta \Delta(\omega)|. 
\end{aligned}
\end{equation}
The intensity variation of the transmitted probe light to a perturbation $\delta\Delta(\omega)$ on the optical resonance is then derived as 
\begin{equation}
\frac{|\delta T(\omega)|}{|\delta \Delta(\omega)|}=\frac{2 \Delta_{0} \kappa_\text{ex} \sqrt{(\kappa-\kappa_\text{ex})^{2}+\omega^{2}}}{\left(\frac{\kappa^{2}}{4}+\Delta_{0}^{2}\right) \sqrt{\left(\frac{\kappa^{2}}{4}+\Delta_{0}^{2}-\omega^{2}\right)^{2}+\omega^{2} \kappa^{2}}}.
\label{sensitivity}
\end{equation}
When the optical microfiber is critically coupled to the microresonator with $\kappa_\text{ex}=\kappa/2$ and the cavity detuning is set at $\Delta_{0}={\kappa}/{2}$, Eq.~\ref{sensitivity} is simplified as $|\delta T(\omega)|/|\delta \Delta(\omega)|=\sqrt{(\kappa^2+4\omega^2)/(\kappa^4+4\omega^4)}$.

Therefore, the optical sensitivity to particle vibrations with frequency $\omega=2\pi \nu$ can be derived as 
\begin{equation}
\begin{aligned}
S_{\mathrm{o}}=|\frac{\delta T(\omega)}{\delta \Delta(\omega)}\frac{\delta (2 \pi f_{\text{o}})}{\delta n_{\text{eff}}}|=&\sqrt{\frac{\kappa^2+4\omega^2}{\kappa^4+4\omega^4}} \cdot \frac{4\pi^2 f_{\text{o}}^2 R_{\text{o}}}{m_{\text{o}}c}  \\
=&\alpha  Q_{\text{o}}  \sqrt{\frac{1+4( \nu Q_{\text{o}} /f_o)^2}{1+4(\nu Q_{\text{o}} /f_{\text{o}})^4}},
\end{aligned}
\end{equation}
where $\alpha=2\pi f_{\text{o}} R_{\text{o}}/(m_{\text{o}}c) $ and $n_{\text{eff}}=m_{\text{o}} c/(2\pi f_{\text{o}}R_{\text{o}})$ with $c$ being the speed of light, $R_{\text{o}}$ the cavity radius, $m_{\text{o}}$ the angular mode number and $Q_{\text{o}} =2\pi f_{\text{o}}/\kappa$ the quality factor of the optical mode. Note that the optical sensitivity of the microresonator can be slightly optimized by adjusting the microresonator-microfiber coupling condition and the frequency detuning between the probe light and the optical mode.

{\bf\noindent III. Simulation analysis of the vibrational modes of microbial cells}

The natural vibrations of cells are analyzed using a core-shell model which accounts for the elasticity of the rigid cell envelop (including cell membrane, cell wall, and outer membrane) and the viscoelastic properties of the soft cytoplasm \cite{zinin2005mechanical} through the finite-element-method (FEM) simulation. The morphologies of cells contacted on the silica microresonator are reproduced and fully meshed according to the microscope images in Extended Data Fig. 3. The mechanical properties of cells are obtained in light of previous studies \cite{liu2019finite}:  the cytoplasm is treated as the incompressible material with Poisson' ratio $\sigma_c=0.49$, the Young’s modulus $E_c$ of less than 1 kPa, the density $\rho_c=$ 1~g/cm$^3$ and the viscosity of $10^{-3}$ Pl; the cell envelop is regarded as a rigid shell with Young's modulus $E_e\sim$ 100 kPa-10 GPa, Passion' ratio $\sigma=0.35$, and 
the density $\rho_e=$1~g/cm$^3$. 

The frequency shifts of the (1,2) quadrupole vibrational modes of the {\it Aspergillus Sydowii} induced by the microwave irradiation are also analyzed. In our experiment, since no apparent change in external morphology of cell is observed under optical microscopy, the large resonance shifts are expected to mainly result from protein denaturation as well as the drying of free water. The heat-induced protein denaturation occurs typically at a specific temperature and could largely increase cell stiffness (\textit{i.e.}, Young's modulus) \cite{woo2000differential, pillet2014uncovering, raikos2007rheology}, while the drying of free water decreases the mass of the cell (\textit{i.e.,} the density of cytoplasma). According to the empirical parameters described above and cell morphology obtained from microscopic images, FEM simulation is performed to analyze three main factors contributed to heating-induced frequency shifts (Figs. S2). It indicates the natural frequency of the (1,2) quadrupole mode is not sensitive to the density and Young's modulus of cytoplasma, while strongly depends on Young's modulus of cell envelop.  
\end{widetext}

\setcounter{figure}{0}
\captionsetup[figure]{labelfont={bf},justification=raggedright,name={Fig. S\hskip 0pt}}
\clearpage
\begin{figure*}[ht]
	\includegraphics[width=18cm]{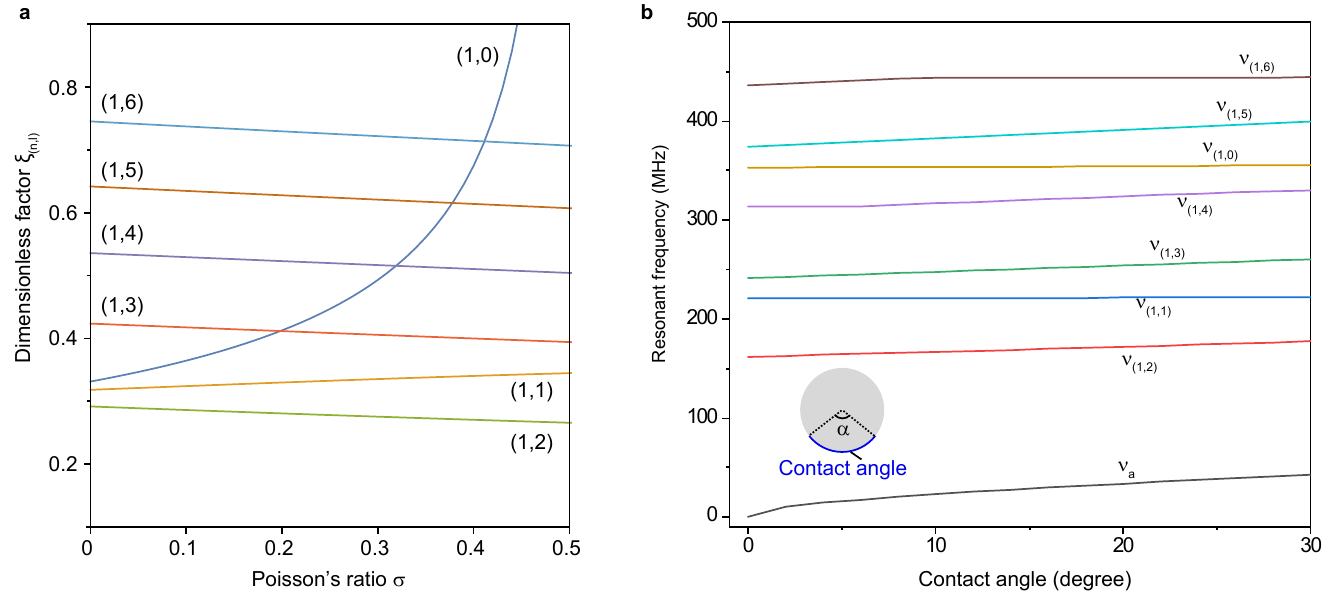}\\
	\caption{\label{s1}\textbf{Natural frequencies and their dependencies on the contact angle with the subtract. a}, The eigenfrequency-relevant dimensionless factor $\xi_{\text (n,l)}$ as a function of Poisson's ratio for the low-order angular vibrational modes $(n=1, l=0\sim6)$ of a free homogeneous spherical particle.   \textbf{b}, Simulated eigenfrequencies of the low-order angular vibrational modes of 2.8-\textmu m-radius polystyrene sphere as a function of contact angle $\alpha$ with the substrate. Inset, the simulation is performed by the finite-element method when a fixed constraint is applied to the boundary with a certain solid contact angle $\alpha$ of the spherical particle.  Generally, $\alpha<10^{\circ}$ for the standard spherical particles in our experiment.
}
\end{figure*}

\begin{figure*}[ht]
\includegraphics[width=15cm]{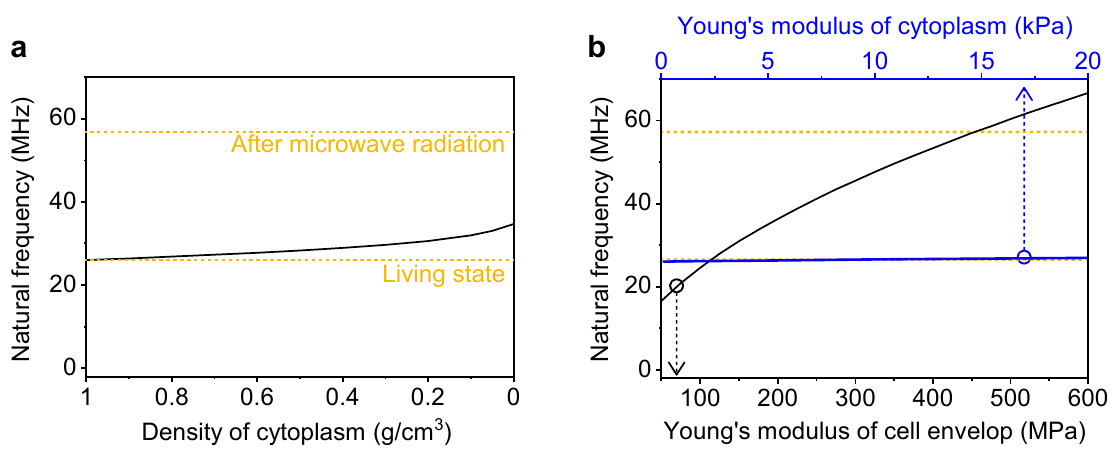}\\
\caption{\textbf{
Analysis of the vibrational frequency of {\textit Aspergillus Sydowii}.} 
{\bf a}, The natural frequency $\nu_{(1,2)}$ of the cell as functions of the density of cytoplasm ({\bf a}). {\bf b}, The natural frequency $\nu_{(1,2)}$ of the cell as functions of Young's moduli of cytoplasm and cell envelop. 
}
\end{figure*}

\begin{figure*}[ht]
\centering
\includegraphics[width=10cm]{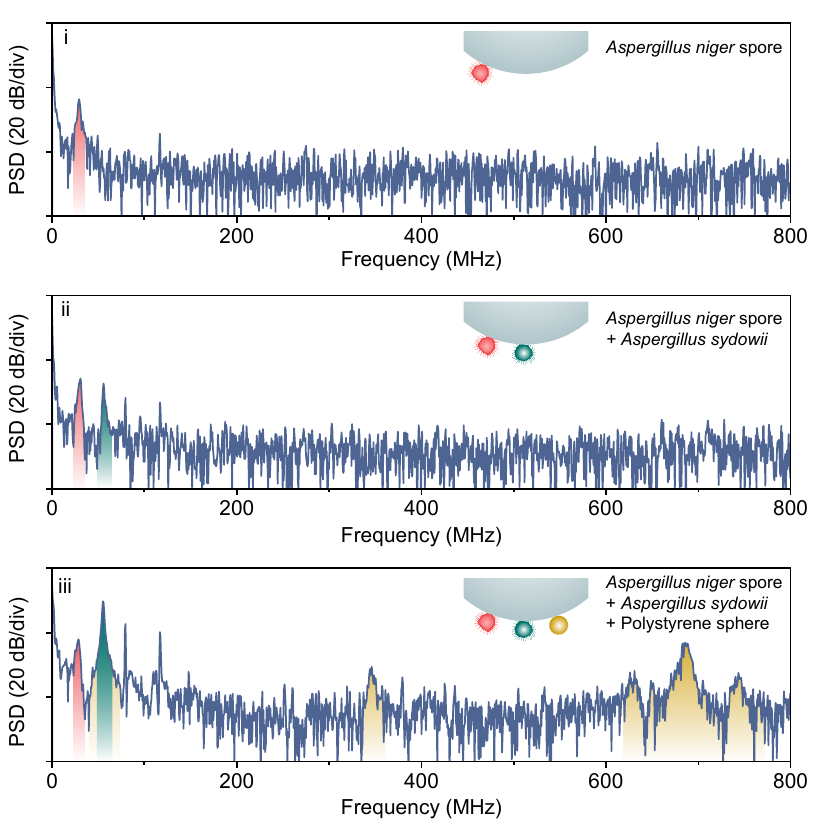}
\caption{\textbf{Mechanical fingerprinting of mixed particles.} Vibrational spectra of mixed particles. i, an {\it Aspergillus niger} spore; ii, an {\it Aspergillus niger} spore and an {\it Aspergillus sydowii}; iii, an {\it Aspergillus niger} spore, an {\it Aspergillus sydowii} and a 1.4-\textmu m-radius polystyrene sphere. The vibrational modes of {\it Aspergillus niger} spore, {\it Aspergillus niger} spore and polystryrene sphere are indicated by red, olive and yellow, respectively.}
\label{Fig5}
\end{figure*}

\end{document}